\documentclass{raa}

\usepackage{url}
\usepackage{subfigure}
\usepackage{longtable}
\usepackage{amsmath}
\usepackage{graphicx,times}
\usepackage{subfigure}
\usepackage{longtable}
\usepackage{txfonts}
\usepackage{lscape}
\usepackage{natbib}

\input{epsf.sty}                        
\begin{document}
   \title {A study of strong pulses detected from PSR B0656+14 using Urumqi 25-m radio telescope at 1540\,MHz}


   \author{Guo-Cun Tao
      \inst{1,2}
   \and Ali Esamdin
      \inst{1}\thanks{Email: aliyi@xao.ac.cn}\
   \and Hui-Dong Hu
      \inst{1,2}
   \and Mao-Fei Qian
      \inst{1,2}
   \and Jing Li
      \inst{1,2}
   \and Na Wang
      \inst{1}      }

   \institute{Xinjiang Astronomical Observatory, Chinese Academy of Sciences, 150 science 1-street,
            Urumuqi, Xinjiang 830011, China
         \and
            Graduate University of Chinese Academy of Sciences, 19A Yuquan Road,
             Beijing 100049, China
             }

   \date{Received~~2012 month day; accepted~~2012~~month day}

\abstract{We report on the properties of strong pulses from PSR
B0656+14 by analyzing the data obtained using Urumqi 25-m radio
telescope at 1540 MHz from August 2007 to September 2010. In 44 hrs
of observational data, a total of 67 pulses with signal-to-noise
ratios above a 5-$\sigma$ threshold were detected. The peak flux
densities of these pulses are 58 to 194 times that of the average
profile, and the pulse energies of them are 3 to 68 times that of
the average pulse. These pulses are clustered around phases about
5\,degrees ahead of the peak of the average profile. Comparing with the
width of the average profile, they are relatively narrow, with the
full widths at half-maximum range from 0.28 to 1.78\,degrees. The
distribution of pulse-energies of the pulses follows a lognormal
distribution. These sporadic strong pulses detected from PSR
B0656+14 are different in character from the typical giant pulses,
and from its regular pulses.
   }

   \authorrunning{G. C. Tao et al. }            
   \titlerunning{A study of strong pulses detected from PSR B0656+14}  

   \maketitle

%
%
\section{Introduction}

The PSR B0656+14 was firstly detected in radio band by
\citet{Manchester+78}. The pulsar has a period of 0.3849s.
The average pulse flux density of the pulsar is 3.7\,mJy at 1.4\,GHz
\citep{Lorimer+95}, and its dispersion measure (DM) is 13.977 pc\,cm$^{-3}$.
The distance inferred from the DM using the free electron density model of
\citet{Cordes+02} is $\approx$ 0.76\,kpc, with a typical error of 20 percent, while
interpretation of X-ray data gives d = 0.2-0.5\,kpc. The average profile
of the pulsar present three components, and the radiation of the main pulse component
is almost completely linearly polarized while two weak components, leading and following the
main, present low-polarization, and are more apparent at low frequencies \citep{Gould+98,Weisberg+04,Hankins+10}.
PSR B0656+14 is an interesting object from which the pulsed emission has been
detected in radio, optical, soft ultraviolet, X-ray, and gamma-ray range of the spectrum.

Recently, very strong individual pulses have been detected from PSR
B0656+14. \citet{Kuzmin+06} detected 52 strong individual pulses
with signal-to-noise ratios (hereafter, SNR) above 5-$\sigma$
detection threshold during approximately 4.8\,hrs of observations at
111\,MHz, and the peak flux of the brightest pulse they noted was up
to 640 times that of the average pulse. \citet{Weltevrede+06b} also
detected strong pulse from the pulsar at 327 MHz, and the highest
peak flux among these strong pulses was 420 times that of the
average pulse. Based on this brightest burst, \citet{Weltevrede+06a}
assumed that if PSR B0656+14 were located at a suitable distance, it
could have been classified as a Rotating Radio Transient.

The individual pulse intensities of most radio pulsars normally
fluctuate several times that of average pulse \citep{Ritchings+76,
Kramer+03}. However, the intensities of typical giant pulses (GPs)
can exceed hundreds and even thousands times that of regular pulses
\citep{Staelin+68,Cairns+04,Knight+05}. The duration of GPs are very
short with timescales down to nano-seconds \citep{Hankins+03}.
Unlike the GPs, the strong pulses detected from PSR B0656+14 are
broad \citep{Weltevrede+06a,Weltevrede+06b}. The essential nature of
the pulsar's strong pulses need to be investigated.

To study the prominent strong pulses of bursts from B0656+14, we
have been monitoring the pulsar at a higher observing frequency of
1540\,MHz since August 2007 . In \S\,2, the observations are
described in detail. In \S\,3, the data analysis procedure and the
results are presented. In \S\,4, we discuss the results. This work
is summarized in \S\,5.


\section{Observations}
The observations were carried out using the Urumqi 25-m radio
telescope. The telescope has a dual-channel cryogenic receiver that
receives orthogonal linear polarizations at the central observing
frequency of 1540\,MHz. The receiver noise temperature is less than
10\,K. Each polarization channel is comprised of 128 sub-channels of
bandwidth 2.5\,MHz, yielding a total bandwidth of 320\,MHz. The data
from each sub-channel are recorded to hard disk with 1-bit sampling
at 0.25\,ms intervals for subsequent off-line processing
\citep{Wang+01}. The minimum detected flux density of 4.8\,Jy at the
5-$\sigma$ threshold is given by
$$
 S_{min} = \frac{2\alpha\beta\kappa T_{sys}}{\eta
A\sqrt{n_p\tau\Delta f}}, \eqno{(1)}
$$
where $\alpha$ = 5 is the SNR, $\beta$ = $\sqrt{\pi/2}$ is a loss
factor due to one-bit digitization, $\kappa$ is the Boltzmann
constant, $T_{sys}$ = $T_{rec}+T_{spl}+T_{sky}\sim$ 32\,K (in which
$T_{rec}$, $T_{spl}$, $T_{sky}$ are the receiver, spillover and the
the sky noise temperatures, respectively), $\eta\approx$ $57\%$ is
the telescope efficiency at 1540\,MHz, $A$ = 490.87\,m$^2$ is the
telescope area, $n_p = 2$ is the number of polarizations channels,
$\tau$ = 0.25\,ms is the sampling interval, and $\Delta f$ =
320\,MHz is the total observing bandwidth of each channel
\citep{Esamdin+08}. The average pulse flux density of PSR B0656+14
is 3.7\,mJy at 1.4\,GHz, so we should note that only very strong single
pulses (if exist) of the pulsar can be detected using the Nanshan 25-m
radio telescope.

From August 2007 to September 2010, 44 hours of data were collected
in 23 observing sessions. For each observation, the sampling
interval was 0.25\,ms, and the time-span lasted about 2\,hrs.

\section{Data Analysis and Results}
\subsection{Data Analysis}
\label{sect:data} Since the group velocities of radio waves are
frequency dependent in the propagation medium, the pulsed radio
radiation from pulsar is dispersed in the ionized plasma of the
interstellar medium. Therefore, the high frequency components of the
radio pulse arrive earlier than those at lower frequencies. These
dispersion delays must be removed in order to obtain the real
profile of a pulse. The time difference $\Delta t$ (in\,ms) between
arrivals of two components at different frequencies is given by

$$
\Delta t = 4.1488 \times DM \times
(\frac{1}{f^{2}_{l}}-\frac{1}{f^{2}_{h}}), \eqno{(2)}
$$

in which DM is the dispersion measure in pc\,cm$^{-3}$, and $f_{l}$
and $f_{h}$ (in GHz) represent the values of the lower and higher
frequencies, respectively. The time delay was calculated using this
equation for each of the 128 observing channels of each
polarization.

The data were de-dispersed by delaying successive channels relative
to the nominal dispersion measure DM = 13.977\,pc\,cm$^{-3}$, of the
pulsar. Then, all pulsed signals above the 5-$\sigma$ threshold were
identified as the pulses candidates. And then, in order to
distinguish instrumental signals or impulsive terrestrial Radio
Frequency Interference (RFI) from the strong pulses of PSR B0656+14,
the de-dispersion procedure was applied from 0.977 to 30.977 \,pc
cm$^{-3}$ at intervals of 0.1 pc\,cm$^{-3}$. The details of the
pulse-identify process see \citet{Esamdin+08}.

A total of 67 pulses were detected through the process mentioned
above from the 44 hours of data. Approximately 1 pulse was detected
for every 6100 rotation periods of the pulsar. Phases of these individual pulses
were calculated by
$$
\Phi (t) =
\Phi_{0}+\nu(t-t_{0})+\frac{1}{2}\dot{\nu}(t-t_{0})^{2}+\frac{1}{6}\ddot{\nu}(t-t_{0})^{3},
\eqno{(3)}
$$
where $\Phi_{0}$=83.23\,degrees is the rotational phase at time $t_{0}$ (MJD 54934.3278), $t$ is the
observation time of the strong individual pulse, $\nu$=2.59796969281\,Hz is the
rotating frequency of the pulsar, $\dot{\nu}$=-3.709653$\times$$10^{-13}$ $s^{-2}$ is the frequency
derivative, $\ddot{\nu}$=1.03$\times$$10^{-24}$ $s^{-3}$ is the frequency second derivative, and all
quantities were defined at the time $t_{0}$ which was obtained by a
observation with average profile of the highest SNR on 13 April
2009.

We investigated the phases of these single pulses by comparing the
residuals of the pulse arrival times obtained using the single-pulse
timing method. In order to compare the phases of the single pulses
with that of the average pulse profile of this pulsar, the average
pulse profile of each observation was generated by folding all
successive individual pulses. Then, we matched the profiles that
were in phase during the complete observation period with the
profile of the highest SNR obtained on 13 April 2009.

\subsection{Results}

\begin{figure}
   \centering
   \includegraphics[angle=0,width=9cm]{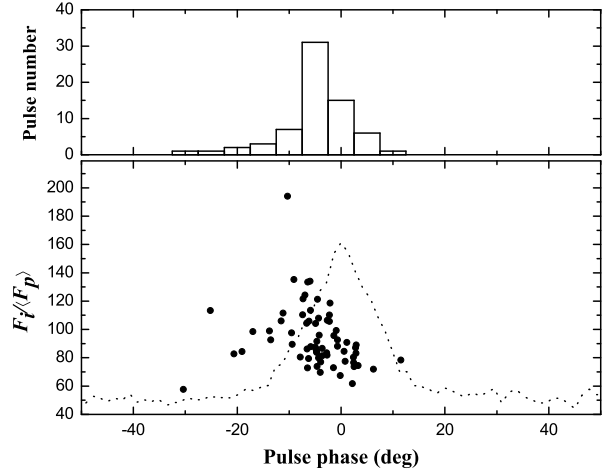}
      \caption{A histogram showing the phase distribution of
      the detected strong pulses (top panel) and the normalized
      peak flux densities versus phases of the strong pulses (bottom panel).
      In order to compare single pulse phases with the phase of the average profile, the average profile are presented by a dotted line in the bottom panel (the y-axis label to the average profile is in an arbitrary unit).}
         \label{FigVibStab}
   \end{figure}

 \begin{figure}
   \centering
   \includegraphics[angle=0,width=10cm]{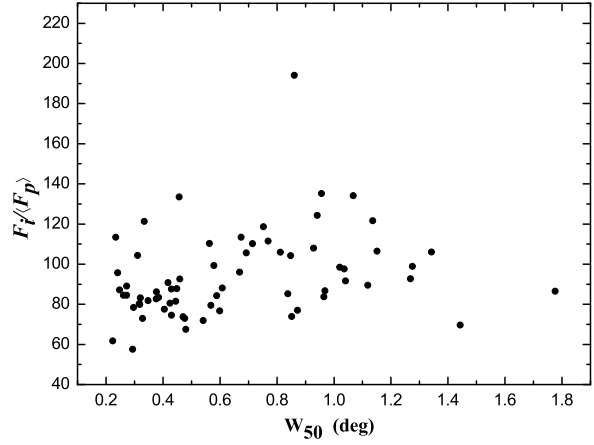}
      \caption{The normalized peak flux densities versus W$_{50}$ of
      the 67 strong pulses detected from PSR B0656+14 at 1540\,MHz.
              }
         \label{FigVibStab}
   \end{figure}

\begin{figure}
   \centering
   \includegraphics[angle=0,width=10cm]{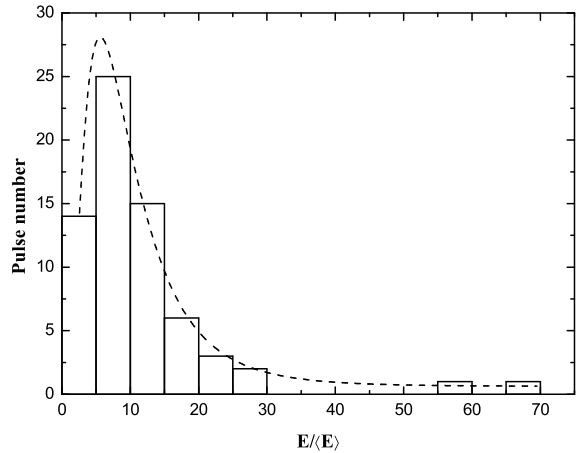}
      \caption{The pulse-energy distribution of the strong pulses
      detected in this work. The energies are normalized to
      the average pulse-energy ($\langle$E$\rangle$) of the pulsar.
      The dashed curve shows a lognormal fitting to the pulse-energy distribution.}
         \label{FigVibStab}
   \end{figure}

The bottom panel of Figure 1 shows the normalized peak flux
densities of the pulses versus their phases (i.e. the peak flux
densities $F_{i}$ of these pulses are normalized to the average peak
flux densities $\langle F_{p} \rangle$). The peak flux densities of
the pulses are above 58 $\langle F_{p} \rangle$, and the brightest
burst is about 194 times stronger than the average pulse. These
pulses are indeed very strong comparing with the average pulse of
the pulsar. We also compare the phases of the strong pulses with the
radiation window of the pulsar by plotting the average profile
(dotted line) in the bottom panel of Figure 1, and where the peak of
the average profile set as the phase of 0 degrees. The top panel of
Figure 1 presents the histogram of the phases of the detected strong
pulses. The strong pulses are distributed over a broad pulse window
with the phase range from -30 to 11.5\,degrees. However, most of the
pulses are clustered around a phase about 5\,degrees earlier than the
phase of the average-profile peak.

Figure 2 shows the normalized peak-flux densities versus full widths
at half maximum (FWHM, W$_{50}$) of the 67 detected pulses. The
W$_{50}$ of these pulses ranged from 0.28 to 1.78\,degrees. Most of the
strong pulses are single-peaked. Comparing with the width of the
average profile, these pulses are narrow.

Figure 3 presents the pulse-energy distribution of the 67 strong
pulses. The pulse energies are normalized to that of the average
profile. Here, the peak of each pulse was arranged to lie in the
512th of 1024 bins. A set of 300 consecutive bins, from the 362nd to
661st, was selected as the pulse window, and 300 consecutive bins of
the remaining bins, which were far away from the peak of each pulse,
were used to remove a baseline from the pulse window. Then, the
energies of each pulse, E, were obtained. The energies of the
average profiles ($\langle$E$\rangle$) were obtained in the same
way. The energies of all single pulses were normalized to the
average pulse-energy ($\langle$E$\rangle$) of the pulsar. The
histogram of pulse energy is presented in Figure 3. The energies of
the pulses range from 3$\langle$E$\rangle$ to 68$\langle$E$\rangle$.

As shown in Figure 3, the pulse-energy distribution is probably best
represented by a lognormal distribution. The equation is given by

$$
{\boldmath P_{longnormal}(E) = P_0 + \frac{A \langle E
\rangle}{\sqrt{2 \pi}\sigma E}exp[-(\ln\frac{E}{\langle E
\rangle}-\mu)^2/(2\sigma^2)]}, \eqno{(4)}
$$
where our best fitting parameters are $P_0$ = 0.62, $A$ = 317.35,
$\mu$ = 2.16 and $\sigma$ = 0.66, respectively.

\section{Discussion}
\label{sect:discussion} By analyzing observation data at\,111MHz,
\citet{Kuzmin+06} suggested that PSR B0656+14 belongs to a group of
pulsars which emit giant pulses. They noted that the pulse-energy
distribution follows a power law, and that those so-called giant
pulses are clustered in a narrow pulse-longitude range
\citep{Kuzmin+06}. The typical GPs are very narrow with timescales
down to nanoseconds \citep{Hankins+03, Knight+06}, and their energy
can easily exceed 10 times that of the average pulse. The GP
phenomenon has only been detected in two young pulsars: Crab pulsar
and PSR B0540-69 \citep{Staelin+68,Wolszczan+84}, and in five
millisecond pulsars
\citep{Romani+01,Johnston+03,Joshi+04,Knight+05}. Although these
seven pulsars have very different rotation rates, all of them have
strong magnetic fields at the light cylinder, B$_{LC}>10^{5}$ gauss.
It is suggested that GPs are inherent in pulsars with extremely
strong magnetic fields at the light cylinder, and that they may
originate near the light cylinder region \citep{Lyutikov+07}.

The magnetic field of PSR B0656+14 at the light cylinder is 766
\,gauss, which is much lower than those of the classical GP emitters (B$_{LC}>10^{5}$ gauss).
The pulse energy of most pulses we detected exceed
10$\langle$E$\rangle$, which are within the range of GP energies.
However, these pulses are much broader than GPs, and distributed
over a wide pulse window. Furthermore, the strong pulses from PSR
B0656+14 showed a lognormal pulse-energy distribution rather than a
power law of GPs. However, due to the small number of strong pulses
we observed, more data are required to confirm whether the
distribution is intrinsically lognormal. Our results are similar to
that presented by \citet{Weltevrede+06a,Weltevrede+06b}. The
sporadic strong pulses of the pulsar are so strong that are hardly
explained as the high end of intrinsical intensity-modulation of
it's regular pulses, and by the interstellar scintillation
considering the timescales of the bursts of pulses and the broad
observing band.

While the intensities of regular single pulses from PSR B0656+14 can
vary at random with values reaching several times the average, are
far below the intensities of observed strong pulses. Furthermore,
the phase distribution of the strong pulses are cluster 5 degree
earlier in longitude than the peak of the average profile of the
pulsar. These may suggest the difference in origin between the
strong pulses detected from the pulsar and its regular pulses.
It may be possible that the sporadic strong pulses and the regular pulses of
PSR B0656+14 represent two different emission modes, i.e. the
strong and weak (normal) modes, with the duration of the strong mode
less than one pulse period. Further studies of the pulsar are necessary.

\section{Summary and Conclusion}
\label{sect:Summary}

We have presented an analysis of the 67 strong pulses detected from
PSR B0656+14 at 1540\,MHz using Urumqi 25 m radio telescope. The
peak flux densities of these pulses are 58 to 194 times of that of
the average pulse, and the pulse energies of them are 3 to 68 times
that of the average pulse. The durations of the strong pulses are
relatively short, ranging from 0.28 to 1.78\,degrees. By covering about
41.5\,degrees of phase range, the strong pulses occurred over a
broad range of the emission window. However, they are mainly
clustered around phase about 5\,degrees earlier than that of the
average-profile peak. The sporadic strong pulses detected from PSR
B0656+14 are different in character from the typical GPs emitted
by some other pulsars, and may also be different in character
from its regular pulses.

\begin{acknowledgements}
     We thank the referee for helpful comments on the paper. This work was funded by the National Natural Science Foundation of China under grant 10973026.
\end{acknowledgements}

\bibliographystyle{raa} 
\bibliography{references} 

\end{document}